\documentclass{aa}  
\usepackage{graphicx}
\usepackage{txfonts}

\def \lsi {LSI+61$^0$303}
\def \kms {km~s$^{-1}$}
\def \rsun {$R_\odot$}
\def \msun {$M_\odot$}
\def \Ha  {$H\alpha$}
\def \vsi {$v \: \sin \: i$}

\begin{document}

   \title{Optical spectroscopy  of Be/gamma-ray binaries}

   \author{R. K. Zamanov\inst{1}
          \and
          K. A. Stoyanov\inst{1}
          \and
          J. Mart\'i\inst{2}
          \and
          G. Y. Latev\inst{1}
          \and 
          Y. M. Nikolov\inst{1}
          \and 
          M. F. Bode\inst{3}
          \and 
          P. L. Luque-Escamilla\inst{4}
          }
   \institute{Institute of Astronomy and National Astronomical Observatory, Bulgarian Academy of Sciences, Tsarigradsko Shose 72, 
        BG-1784 Sofia, Bulgaria
         \and
       Departamento de F\'isica (EPSJ), Universidad de Ja\'en, Campus Las Lagunillas,  A3-420, 23071, Ja\'en, Spain
          \and
        Astrophysics Research Institute, Liverpool John Moores University, IC2 Liverpool Science Park, Liverpool, L3 5RF, UK
          \and
       Departamento de Ingenier\'ia  Mec\'anica y Minera (EPSJ), Universidad de Ja\'en, Campus Las Lagunillas A3-008, 23071, Ja\'en, Spain 
          }

   \date{Received April 20, 2016; accepted  July 8, 2016}

 \abstract{ We report optical spectroscopic observations of the Be/$\gamma$-ray binaries LSI+61303,  MWC~148 and MWC~656.
  The peak separation and equivalent widths of prominent emission lines 
  (\Ha, H$\beta$, H$\gamma$, HeI, and FeII) are measured. 
  We estimated the circumstellar disc size, 
  compared it with separation between the components, and discussed the disc truncation. 
  We find that  in \lsi\ the compact object comes into contact with the outer parts of the circumstellar disc at periastron,  
  in MWC~148 the compact object goes deeply into the disc during the periastron passage, and
  in MWC~656 the black hole is  accreting from the outer parts of the circumstellar disc along the entire orbit. 
  
  The interstellar extinction was estimated using interstellar lines. The rotation
  of the mass donors appears to be  similar to the rotation of the mass donors in Be/X-ray binaries. 
  We suggest that X-ray/optical  periodicity $\sim \! 1$ day deserves to be searched for. }
               
   \maketitle
%

\section{Introduction}
The rapid and sustained progress of high energy and very high energy  astrophysics in recent years enabled
the identification of a new group of binary stars emitting at TeV energies
(e.g. Paredes et al. 2013). These objects, called
$\gamma$-ray binaries, are high-mass X-ray binaries that consist of a compact object (neutron star or black hole)  
orbiting an optical companion that is an OB star. 
There are five confirmed $\gamma$-ray binaries so far: PSR~B1259-63/LS 2883 (Aharonian et al. 2005), 
LS~5039/V479~Sct (Aharonian et al. 2006), \lsi\ (Albert et al., 2009), HESS~J0632+057/MWC~148 (Aharonian et al. 2007), 
and 1FGL~J1018.6-5856 (H.E.S.S. Collaboration et al. 2015). 
Their most distinctive fingerprint is a spectral energy distribution dominated by non-thermal photons with energies up to the TeV domain.
Recently, Eger et al. (2016) proposed a binary nature for the $\gamma$-ray source
HESS~J1832-093/2MASS~J18324516-092154
 and this object probably belongs to the family of the $\gamma$-ray binaries as a sixth member.

The binary system, PSR B1259-63 is unique, since it is the only one where the compact object has been identified
as a radio pulsar (Johnston et al. 1992, 1994). 
The nature of the compact object is known in  PSR B1259-63 as a neutron star, and in AGL~J2241+4454/MWC~656 
as a black hole (Casares et al. 2014). 
Although not included in the confirmed list, MWC~656 was selected as a target here 
despite not having shown all the observational properties of a canonical $\gamma$-ray binary yet.
It was only occasionally detected by the AGILE observatory at GeV energies and not yet detected in the TeV domain (see 
Aleksi{\'c} et al. 2015). 
Nevertheless, the fact that the black hole nature of the compact companion is 
almost certain renders it very similar to the typical $\gamma$-ray binaries. 
In the other systems the nature of the compact object remains unclear  (e.g.  Dubus 2013). 
In addition to these objects, there are several other binary systems 
($\eta$~Car, Cyg~X-1, Cyg~X-3, Cen~X-3, and SS~433) 
that are detected as GeV sources, but not as TeV sources so far.

Here we report high-resolution spectral observations of  \lsi,   MWC~148, and MWC~656,  
and discuss circumstellar disc size, disc truncation, interstellar extinction, and rotation of their mass donors. 
The mass donors (primaries) of these three targets  are emission-line Be stars. 
The Be stars are non-supergiant, fast-rotating B-type and luminosity class III-V stars which, 
at some point in their lives, have shown spectral lines in emission (Porter \& Rivinius 2003). 
The material expelled from the equatorial belt of a 
rapidly rotating Be star forms an outwardly diffusing gaseous, dust-free Keplerian disc 
(Rivinius et al. 2013).    
In the optical/infrared band, 
the two most significant observational characteristics of Be stars  and their excretion discs
are  the emission lines and the infrared excess. 
Moving along the orbit, the compact object passes close to this disc, and sometimes may even go through it 
causing significant perturbations in its structure. This circumstellar disc feeds
the accretion disc around the compact object  and/or interacts with its relativistic wind.

\section{Observations}

\begin{table}
\caption{Journal of observations}             
\label{tab.J}      
\centering          
\begin{tabular}{c c c c l l l }     
\hline\hline       
Date-obs          & exp-time &    S/N       &  Orb. phase & \\ 
yyyymmdd...hhmm  &          &  $H\alpha$   &             & \\
\hline     
\\               
{\bf   \lsi\  }  &  \\
20140217...1923 & 60 min & 20 & 0.455 & \\
20140314...1746 & 60 min & 42 & 0.396 & \\
20150805...0009 & 60 min & 45 & 0.579 & \\
\\       
{\bf  MWC~148 }  &  \\
20140113...1857 & 60 min & 56 &  0.758 &  \\
20140217...2031 & 60 min & 44 &  0.870 &  \\
20140218...1826 & 60 min & 62 &  0.872 &  \\
20140313...2002 & 60 min & 54 &  0.946 &  \\
20140314...1855 & 60 min & 81 &  0.949 &  \\
20140315...1833 & 60 min & 46 &  0.952 &  \\              
\\
{\bf  MWC~656 } &  \\
20150705...2259 & 30 min & 55 &  0.691 &  \\ 
20150804...0017 & 30 min & 45 &  0.173 &  \\
20150804...2229 & 30 min & 56 &  0.188 &  \\
\\
\hline                  
\end{tabular}
\end{table}

High-resolution optical spectra of the three northern Be/$\gamma$-ray binaries were secured with the 
fibre-fed Echelle spectrograph {\it ESpeRo}
attached to the 2.0 m telescope of the National Astronomical Observatory Rozhen, 
located in Rhodope mountains, Bulgaria.
The spectrograph uses  R2 grating with 37.5 grooves/mm,
Andor CCD  camera  2048 x 2048 px, 13.5x13.5 $\mu m$~px$^{-1}$ (Bonev et al. 2016). 
The spectrograph provides a dispersion of 0.06 \AA\,px$^{-1}$ at 6560 \AA\   
and  0.04 \AA\,px$^{-1}$  at 4800 \AA. 

The spectra were reduced in the standard way including bias removal, flat-field correction, 
and wavelength calibration. 
Pre-processing of data and parameter measurements are performed using various routines provided in IRAF. 
The journal of observations is presented  in Table~\ref{tab.J}, where 
the date, start of the exposure, exposure time, and signal-to-noise ratio at about $\lambda 6600$~\AA\ are given.
The orbital phases are calculated 
using $HJD_0= 2443366.775$, $HJD_0= 2454857.5,$ and $HJD_0= 2453243.7$ for \lsi, MWC~148, and MWC~656, respectively, 
and orbital periods given in Sect.~\ref{sect.2}. 

\begin{figure*}   
  \vspace{11.0cm}   
  \includegraphics{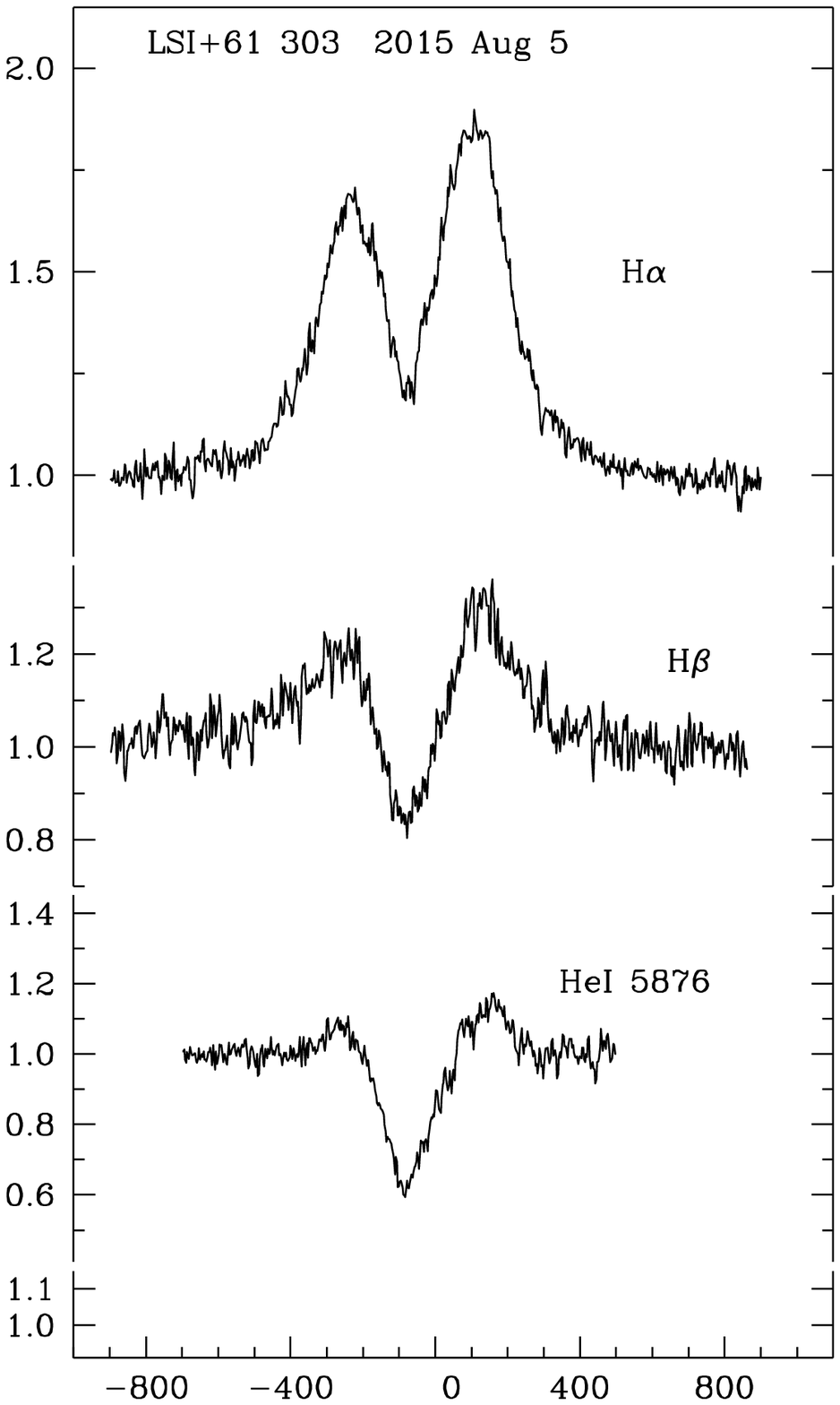}      
  \includegraphics{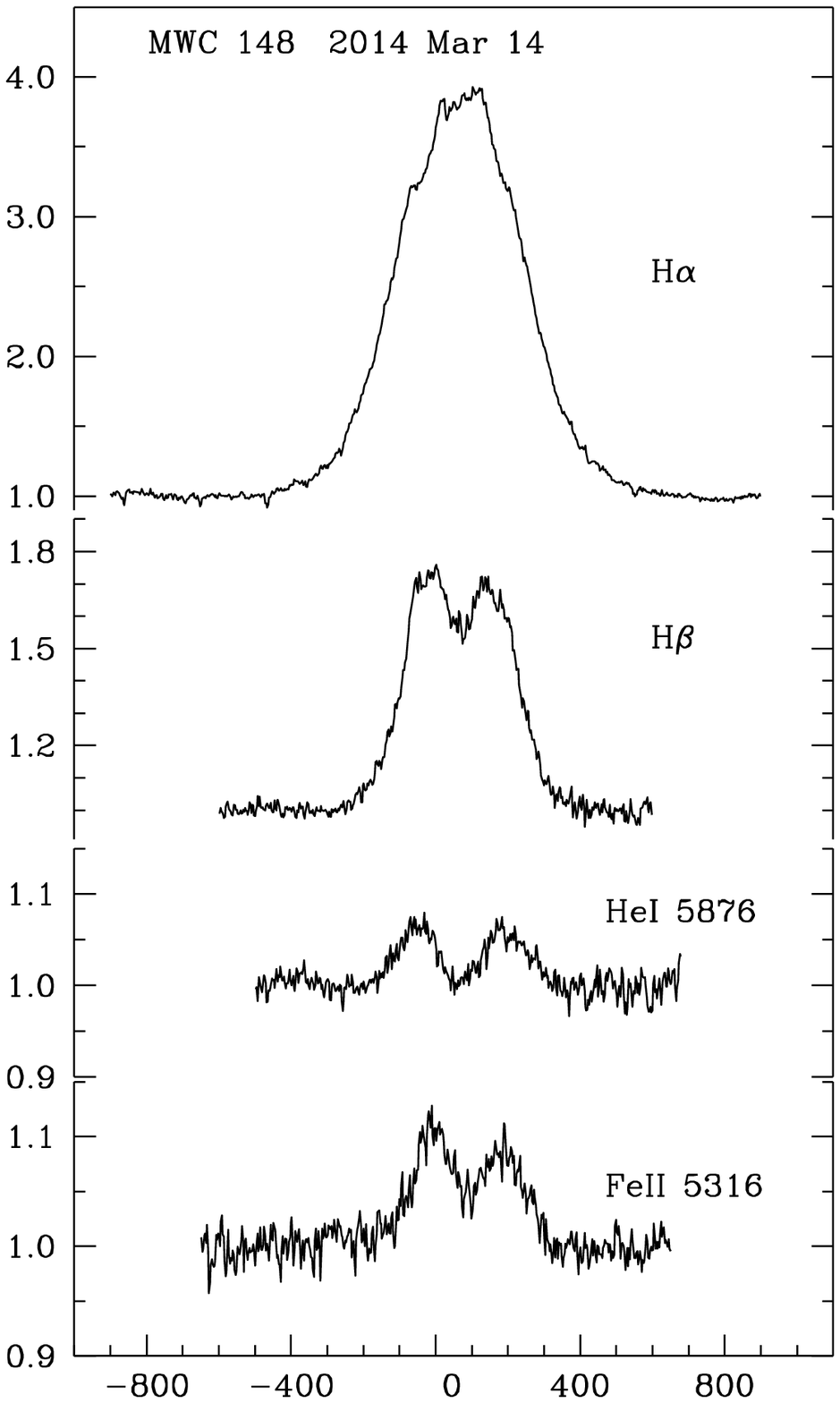}      
  \includegraphics{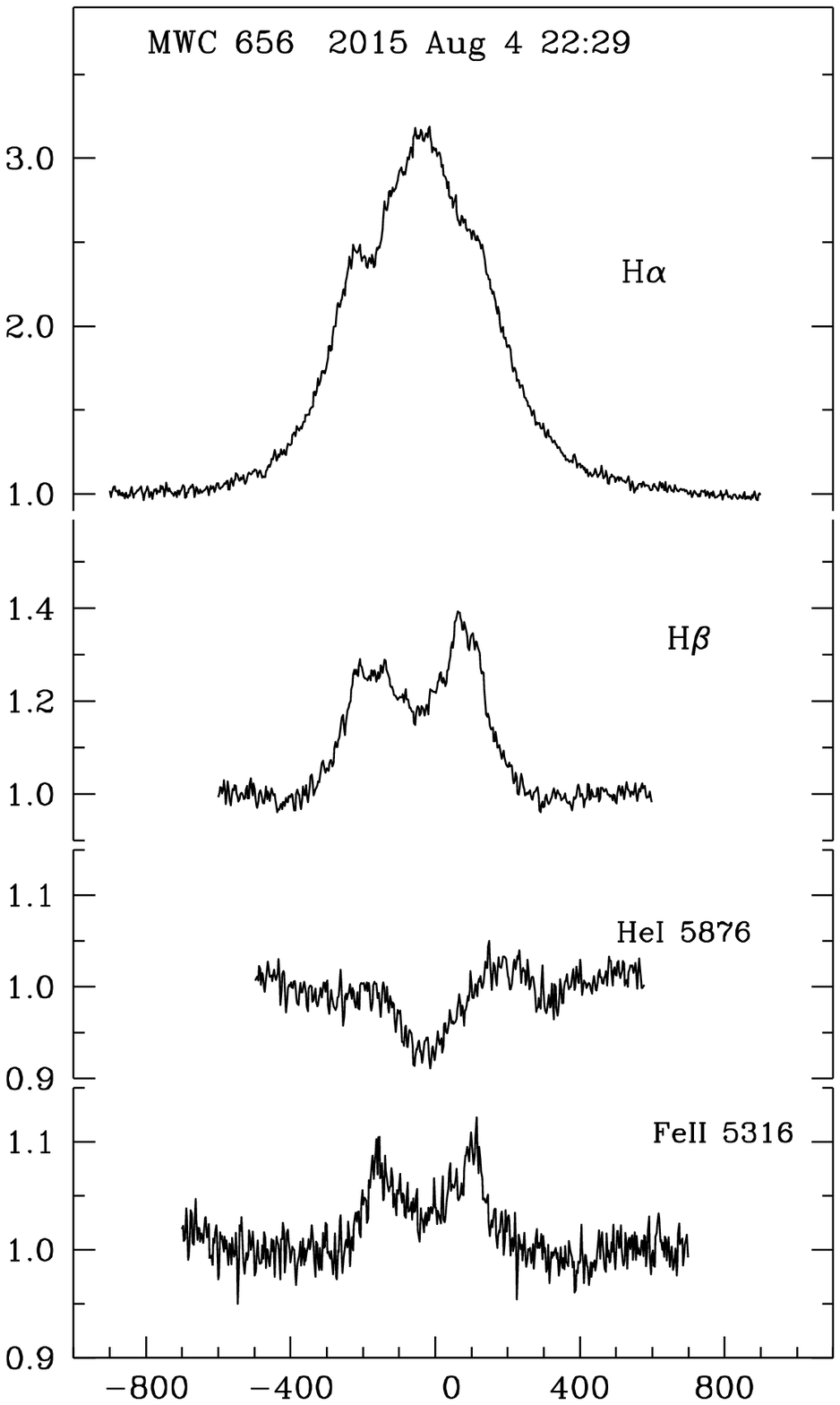}      
\caption[]{Emission line profiles of \lsi, MWC~148, and MWC~656. }  
\label{f1.examp}      
\end{figure*}

Emission line profiles of \lsi, MWC~148, and  MWC~656 are plotted on Fig.\ref{f1.examp}.
Spectral line parameters equivalent width (W) and distance between the peaks ($\Delta V$) for 
the prominent lines ($H\alpha$, H$\beta$, H$\gamma$, $HeI \lambda 5876,$  and $FeII  \lambda 5316$) 
are given in Table~\ref{tab.2}. 
The typical error on the equivalent width is below $\pm 10$~\%   for  
lines with $W > 1$~\AA\  and up to $\pm 20$\% for lines with $W \lesssim  1$~\AA. 
The typical error on $\Delta V$ is $\pm 10$~\kms.  
It is worth noting that
{\bf (1)} in \lsi\  FeII lines are not detectable; 
{\bf (2)} In MWC~656 on spectrum  20150705 the HeI~$\lambda5876$ line is not visible (probably emission 
fills up the absorption). 

In addition to the Rozhen data we use  98 spectra of MWC 148 and 68 spectra of MWC 656
(analysed in Casares et al. 2012)  
from the archive of the 2.0 m Liverpool Telescope\footnote{The Liverpool Telescope is operated 
on the island of La Palma by Liverpool John Moores University 
in the Spanish Observatorio del Roque de los Muchachos of the Instituto de Astrofisica de Canarias 
with financial support from the UK Science and Technology Facilities Council.}  (Steele et al. 2004).
These spectra were obtained using the Fibre-fed RObotic Dual-beam Optical Spectrograph (FRODOSpec; Morales-Rueda et al. 2004).
The spectrograph is fed by a fibre bundle array consisting of $12\times12$ lenslets of 0.82 arcsec each,
which is reformatted as a slit.
The spectrograph was operated in a high-resolution mode, providing a dispersion of
0.8 \AA\,px$^{-1}$ at 6500 \AA,  0.35 \AA\,px$^{-1}$ at 4800 \AA, and typical $S/N \gtrsim 100$. 
FRODOSpec spectra were processed using the fully automated data reduction pipeline of Barnsley et al. (2012).
The typical error on the equivalent width is  $\pm 10$~\%  and on $\Delta V$ is $\pm 20$~\kms. 

\begin{table*}
\caption{Spectral line parameters}                  
\centering          
\begin{tabular}{cccccccccccccccccccclll}   
\hline\hline 
{\bf object}   & \multicolumn{2}{c}{H$\alpha$}  & \multicolumn{2}{c}{H$\beta$} & \multicolumn{2}{c}{H$\gamma$}  & \multicolumn{2}{c}{HeI 5876} & \multicolumn{2}{c}{FeII 5316}  &    \\  
date-obs       & $W_\alpha$ & $\Delta V_\alpha$ & $W_\beta$   & $\Delta V_\beta$ &  $W_\gamma$  & $\Delta V_\gamma$  &  $W_{HeI5876}$   &  $\Delta V_{HeI5876}$ &  $W_{FeII5316}$    & $\Delta V_{FeII}$ &  \\   
yyyymmdd.hhmm  & \AA\ &   \kms\    & \AA\ &     \kms      & \AA\ &   \kms\    & \AA\  &    \kms\   & \AA\  &    \kms\   &   \\ 
\hline     
\\               
{\bf   \lsi\  }  &  \\
20140217.1923 & -8.6 & 316 & -0.71  & 416 & +0.5 & ... & +0.39 & 455 & ... &  ... & \\ 
20140314.1746 & -8.1 & 309 & -1.12  & 411 & +0.7 & ... & +0.17 & 456 & ... &  ... & \\
20150805.0009 & -8.2 & 337 & -1.16  & 421 & +0.8 & ... & +0.40 & 416 & ... &  ... & \\
\\
{\bf  MWC~148 } &  \\
20140113.1857 & -29.5 & 105 & -4.19 & 182 & -1.38 & 188 & -0.49 & 231 & -0.49 & 210 &  \\
20140217.2031 & -30.9 &  92 & -4.27 & 162 & -1.24 & 178 & -0.37 & 243 & -0.37 & 177 &  \\
20140218.1826 & -29.3 &  87 & -4.12 & 161 & -1.12 & 171 & -0.39 & 244 & -0.39 & 182 &  \\
20140313.2002 & -29.0 & ... & -3.41 & 165 & -0.77 & 149 & -0.39 & 246 & -0.39 & 186 &    \\
20140314.1855 & -28.5 & ... & -3.79 & 170 & -1.04 & 159 & -0.37 & 251 & -0.37 & 185 &    \\
20140315.1833 & -26.8 & ... & -3.84 & 168 & -1.01 & 175 & -0.51 & 264 & -0.51 & 186 &    \\ 
\\
{\bf  MWC~656 } &  \\
20150705.2259 & -23.3 & ... & -2.26 & 246 & -0.42 & 301 & +0.0  & ...  & -0.40 & 275 &   \\ 
20150804.0017 & -21.9 & ... & -2.12 & 244 & -0.23 & 289 & +0.22 & ...  & -0.48 & 240 &   \\
20150804.2229 & -21.2 & ... & -1.98 & 246 & -0.34 & 311 & +0.19 & ...  & -0.42 & 227 &   \\
\\
\hline                  
\end{tabular}
\label{tab.2}   
\end{table*}

\section{Objects: System parameters}
\label{sect.2}

\lsi\  (V615~Cas) was identified as a $\gamma$-ray source with the $COS B$ satellite  35 years ago (Swanenburg et al. 1981). 
For the orbital period of 
\lsi,  we adopt  $P_{orb}=26.4960 \pm 0.0028$~d,   
which was derived with Bayesian analysis of radio observations (Gregory 2002) 
and an orbital eccentricity $e=0.537$, which was obtained on the basis of the radial velocity of the primary 
(Casares et al. 2005; Aragona et al. 2009).    
For the primary, Grundstrom et al. (2007) suggested a B0V star 
with radius $R_1=6.7 \pm 0.9$~\rsun. 
A B0V star is expected to have on average $M_1 \approx 15$~\msun\ (Hohle et al. 2010). 
  We adopt  
$v \sin i = 349 \pm 6$~\kms\  for the projected rotational velocity of the mass donor (Hutchings \& Crampton 1981, Zamanov et al. 2013).  

MWC~148  (HD 259440)   was identified as the counterpart of the variable TeV source HESS J0632+057 (Aharonian et al. 2007). 
We adopt  $P_{orb} = 315 ^{+6}_{-4}$~d derived from the X-ray data (Aliu et al. 2014), which is consistent with the 
previous result of $321 \pm 5$ days (Bongiorno et al.  2011). 
For this object  Aragona et al. (2010)
derived  $T_{eff} = 27500 - 30000$~K,  $\log g = 3.75 - 4.00$, $M_1 = 13.2 - 19.0$~\msun, and  $R_1 = 7.8 \pm 1.8 $~\rsun. 
For the calculations in Sect.\ref{Disc.size}, 
we adopt  $e=0.83$, periastron at phase 0.967 (Casares et al. 2012), and $v \sin i = 230 - 240$~\kms\   (Moritani et al. 2015).

MWC~656 (HD 215227) is the emission-line Be star that lies within the positional error circle 
of the $AGILE$ $\gamma$-ray source  AGL J2241+4454 (Lucarelli et al. 2010).
It is the first and until now the only detected binary composed of a Be star 
and a black hole (Casares et al. 2014).  
For the orbital period, 
we adopt   $P_{orb}=60.37 \pm 0.04$~d obtained with optical photometry  (Williams et al. 2010), 
$e=0.10 \pm 0.04$ estimated on the basis of the radial velocity measurements 
and $v \sin i = 330 \pm 30$~\kms\ (Casares et al. 2014).
For the primary, Williams et al. (2010) estimated $T_{eff} = 19000 \pm 3000$~K, $\log g = 3.7 \pm 0.2 $, 
$M_1 = 7.7 \pm 2.0$~\msun,  $R_1 = 6.6 \pm 1.9$~\rsun. 
Casares et al. (2014) considered that the mass donor is a giant  (B1.5-2~III) 
and give a mass range  $M_1 = 10 - 16$~\msun. 
On average a B1.5-2~III star is expected to have about $R_1 \approx 8.3 - 8.8$~\rsun\  (Straizys \& Kuriliene 1981). 
From newer values of the luminosity (Hohle et al. 2010), such
a star is expected  
to have  $M_1 \approx 8.0 - 10.0$~\msun\ and radius $R_1 \approx  9.5 - 10$ \rsun.
We adopt $R_1 \approx 10$ \rsun\  for the calculations in Sect.\ref{Disc.size}. 

\begin{table*}
\caption{Disc size in different emission lines}                 
\centering          
\begin{tabular}{cccc|cccccllllll}      
\hline\hline       
Date-obs  & $R_{disc}(H\alpha)$ &   $R_{disc}(H\alpha)$ & $R_{disc}(H\alpha)$ & $R_{disc}(H\beta)$ & $R_{disc}(H\gamma)$ & $R_{disc}(HeI5876)$ & $R_{disc}(FeII)$ & \\ 
yyyymmdd.hhmm &    \rsun\           &    \rsun\         &   \rsun\            &     \rsun\          &    \rsun\           &       \rsun\        &      \rsun\      & \\
          & (\tablefootmark{a}) &   (\tablefootmark{b}) & (\tablefootmark{c}) &                    &                     &                     &                  & \\
\hline     
\\  
{\bf   \lsi\  } & \\
20140217.1923 &  33\tablefootmark{a} & 32\tablefootmark{b} & 36\tablefootmark{c}  & 19  & ... &  16 & ... &  \\
20140314.1746 &  34\tablefootmark{a} & 33\tablefootmark{b} & 33\tablefootmark{c}  & 19  & ... &  16 & ... &  \\
20150805.0009 &  29\tablefootmark{a} & 31\tablefootmark{b} & 34\tablefootmark{c}  & 18  & ... &  19 & ... &  \\
\\
{\bf  MWC~148 } & \\
20140113.1857 & 156\tablefootmark{a} & 165\tablefootmark{b} & 180\tablefootmark{c} & 52 & 49 & 32 & 39 &  & \\
20140217.2031 & 205\tablefootmark{a} & 208\tablefootmark{b} & 190\tablefootmark{c} & 66 & 54 & 29 & 55 &  & \\
20140218.1826 & 226\tablefootmark{a} & 211\tablefootmark{b} & 178\tablefootmark{c} & 66 & 59 & 29 & 52 &  & \\
20140313.2002 & ...                  & 201\tablefootmark{b} & 176\tablefootmark{c} & 63 & 52 & 29 & 50 &  & \\
20140314.1855 & ...                  & 189\tablefootmark{b} & 172\tablefootmark{c} & 60 & 56 & 27 & 50 &  & \\
20140315.1833 & ...                  & 193\tablefootmark{b} & 160\tablefootmark{c} & 60 & 50 & 25 & 50 &  & \\   
\\
{\bf  MWC~656 } & \\
20150705.2259 & ...                  & 213\tablefootmark{b} & 174\tablefootmark{c} &  63 &  42 & ... & 51 &  &  \\ 
20150804.0017 & ...                  & 216\tablefootmark{b} & 162\tablefootmark{c} &  64 &  46 & ... & 66 &  &  \\
20150804.2229 & ...                  & 213\tablefootmark{b} & 156\tablefootmark{c} &  63 &  40 & ... & 74 &  &  \\
\\
\hline  
\label{tab.D}               
\end{tabular}
\tablefoot{
\tablefoottext{a}{Calculated using $\Delta V_\alpha$ and Eq.~\ref{Huang}. }
\tablefoottext{b}{Calculated using $\Delta V_\beta$  and Eq.~\ref{Huang}. }
\tablefoottext{c}{Calculated using $W_\alpha$  and Eq.~\ref{Rd.W2}.}
}
\end{table*}

\section{Circumstellar disc}

\subsection{Peak separation in different lines}
\label{peak.sep}
For the Be stars, the peak separations in  different lines follow approximately 
the  relations (Hanuschik et al. 1988)

\begin{eqnarray}      
 \Delta V_\beta      \approx 1.8 \Delta V_\alpha                                   \label{H3.1}    \\
 \Delta V_\gamma     \approx 1.2 \Delta V_\beta \approx 2.2 \Delta V_\alpha        \label{H3.2}    \\
 \Delta V_{\rm FeII} \approx 2.0 \Delta V_\alpha                                   \label{H3.3}    \\ 
 \Delta V_{\rm FeII} \approx 1.1 \Delta V_\beta                                    \label{H3.4}    
,\end{eqnarray}
where Eq. \ref{H3.4} is derived from Eqs. \ref{H3.1} and \ref{H3.3}.

For \lsi\ using the measurements in Table~\ref{tab.2},  we obtain  
$\Delta V_\beta = 1.30 \pm 0.04 \, \Delta V_\alpha$ and 
$\Delta V_{HeI5876}= 1.38 \pm 0.13 \, \Delta V_\alpha $. The ratio $\Delta V_\beta / \Delta V_\alpha$ is 
considerably below the average value for the Be stars (see Eq.\ref{H3.1}). 

We obtain 
$\Delta V_\beta   = 1.78 \pm 0.06 \, \Delta V_\alpha$,   
$\Delta V_\gamma  = 1.07 \pm 0.03 \, \Delta V_\beta$, and  
$\Delta V_{FeII5316} = 1.12 \pm 0.03 \, \Delta V_\beta$,
$\Delta V_{HeI5876}  = 1.47 \pm 0.10 \, \Delta V_\beta$ for MWC~148.
We use only three spectra for $H\alpha$  (20140113, 20140217, and 20140218) when two peaks in \Ha\ are 
visible. 
The value of  $\Delta V_\beta  / \Delta V_\alpha \approx 1.78$   is very similar to 1.8 in Be stars,
the ratio  $\Delta V_{FeII5316} / \Delta V_\beta \approx 1.07 $ is similar to 1.1 in Be stars,   
and the value of  $\Delta V_\gamma / \Delta V_\beta \approx 1.07$ is again similar to the value 1.2 for Be stars.

We estimate 
$\Delta V_\beta   = 1.72 \pm 0.18 \, \Delta V_\alpha$ 
for MWC~656 (using  six spectra from the Liverpool Telescope  FRODOSpec, where two peaks are visible 
in both $H\alpha$ and $H\beta$), 
$\Delta V_\gamma  = 1.22 \pm 0.04 \, \Delta V_\beta$,
$\Delta V_{FeII5316} = 1.01 \pm 0.10 \, \Delta V_\beta$,  
where all three ratios 
are  similar to the corresponding values (Eq. \ref{H3.1},  \ref{H3.2}, \ref{H3.4}) in Be stars. 
We do not see two peaks on high-resolution Rozhen spectra of this object. 
However two peaks are clearly distinguishable on a few of the LT spectra. 
In every case of detection/non-detection of the double peak structure, 
$W\alpha$ is very similar $19 < W\alpha < 25$~\AA.

The comparison of the peak separation of different emission lines shows 
that MWC~148 and MWC~656 have circumstellar disc that is similar to that of the normal Be stars.
At this stage considerable deviation from the behaviour of the  Be stars is only detected in  \lsi.
In this star the $H\alpha$-emitting disc is only  1.7  times larger than the H$\beta$-emitting disc, 
while in normal Be stars it is 3.3 times larger. This probably is one more indication that
outer parts of the disc are truncated as a result of the relatively short orbital period.   

 \begin{figure}   
  \vspace{9.0cm}   
  \includegraphics{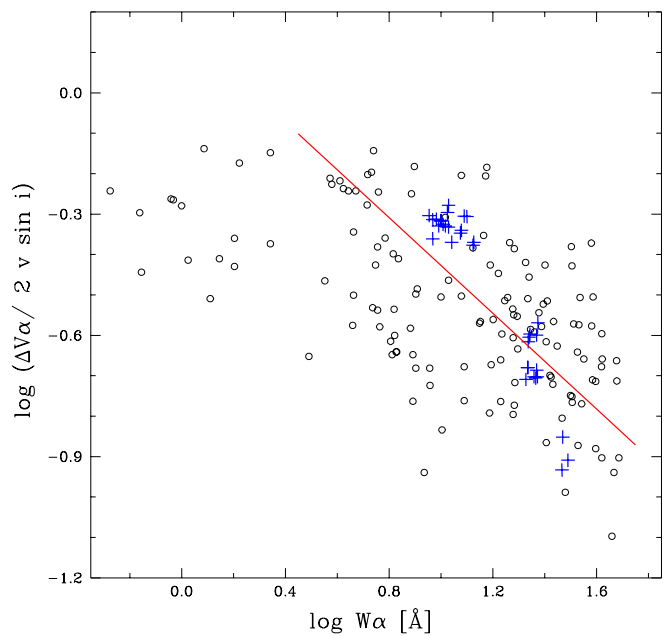}      
  \caption[]{Distance between the peaks of  $H\alpha$ emission line 
  (normalized with stellar rotation and inclination)
   versus $W_\alpha$ on a logarithmic scale. 
   Black empty circles indicate normal Be stars,  
   red pluses indicate $Be/\gamma$-ray binaries. 
   The solid line denotes $y =- 0.592 \: x + 0.165 $ (see Sect.~\ref{RdWa}).   }  
\label{f3.EW.dV}      
\end{figure}         

\subsection{Disc size}
\label{Disc.size}

For rotationally dominated profiles, the peak separation can be regarded as a measure of 
the outer radius ($R_{disc}$) of the emitting disc (Huang 1972)  
 \begin{equation}
      \left( \frac{\Delta V}
                {2\,v\,\sin{i}} \;\right)
       = \;  \left( \frac {R_{disc}}{R_1}\;\right)^{-j} ,
  \label{Huang}
  \end{equation}
where $j=0.5$ for Keplerian rotation, $j=1$ for angular momentum conservation,
$R_1$ is the radius of the primary, and $v\,\sin{i}$ is its projected rotational velocity. 
Eq. \ref{Huang} relies on the assumptions that (1) the Be star is rotating critically, 
and (2) that the line profile shape is dominated by kinematics and radiative transfer does not play a role. 

When the two peaks are visible in the emission lines, we can estimate the disc radius using Eq.~\ref{Huang}.
The calculated disc size for different emission lines are given in Table~\ref{tab.D}. 

In the $H\alpha$  emission line of  \lsi\  two peaks are clearly visible on all of our spectra. 
However in  MWC~656 the $H\alpha$ emission line  seems to exhibit three peaks 
(see Fig.~\ref{f1.examp}).  
Two peaks in $H\alpha$ emission of MWC~148 are clearly detectable on January-February 2014 observations. 
Two peaks are not distinguishable on the spectra obtained in March 2014 (when the companion is at periastron),  
which probably indicates perturbations in the outer parts of the disc caused by 
the orbital motion of the compact object. 

In the $H\beta$ line two peaks are  visible on all the Rozhen spectra. 
We take this opportunity to obtain an estimation of the $R_{disc}(H\alpha)$
using  $\Delta V_\beta$;  the ratios  $\Delta V_\beta / \Delta V_\alpha$  (as obtained 
in Sect.~\ref{peak.sep}), and Eq.\ref{Huang}. 
The $R_{disc}(H\alpha)$ values calculated in this way are given in Table~\ref{tab.D} and indicated with $(^b)$.

 \begin{figure*}     
  \vspace{7.0cm}    
  \includegraphics{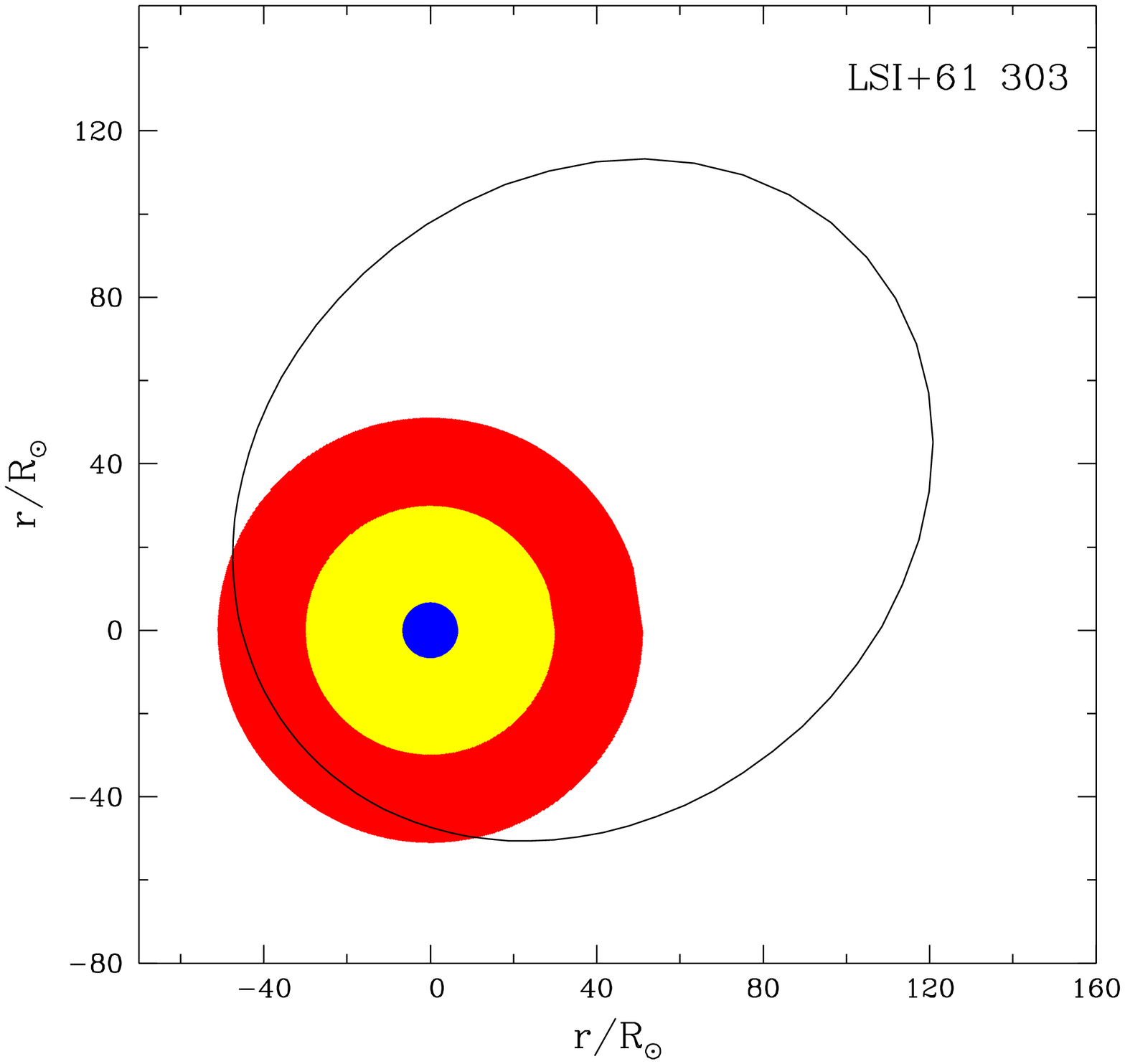}  
  \includegraphics{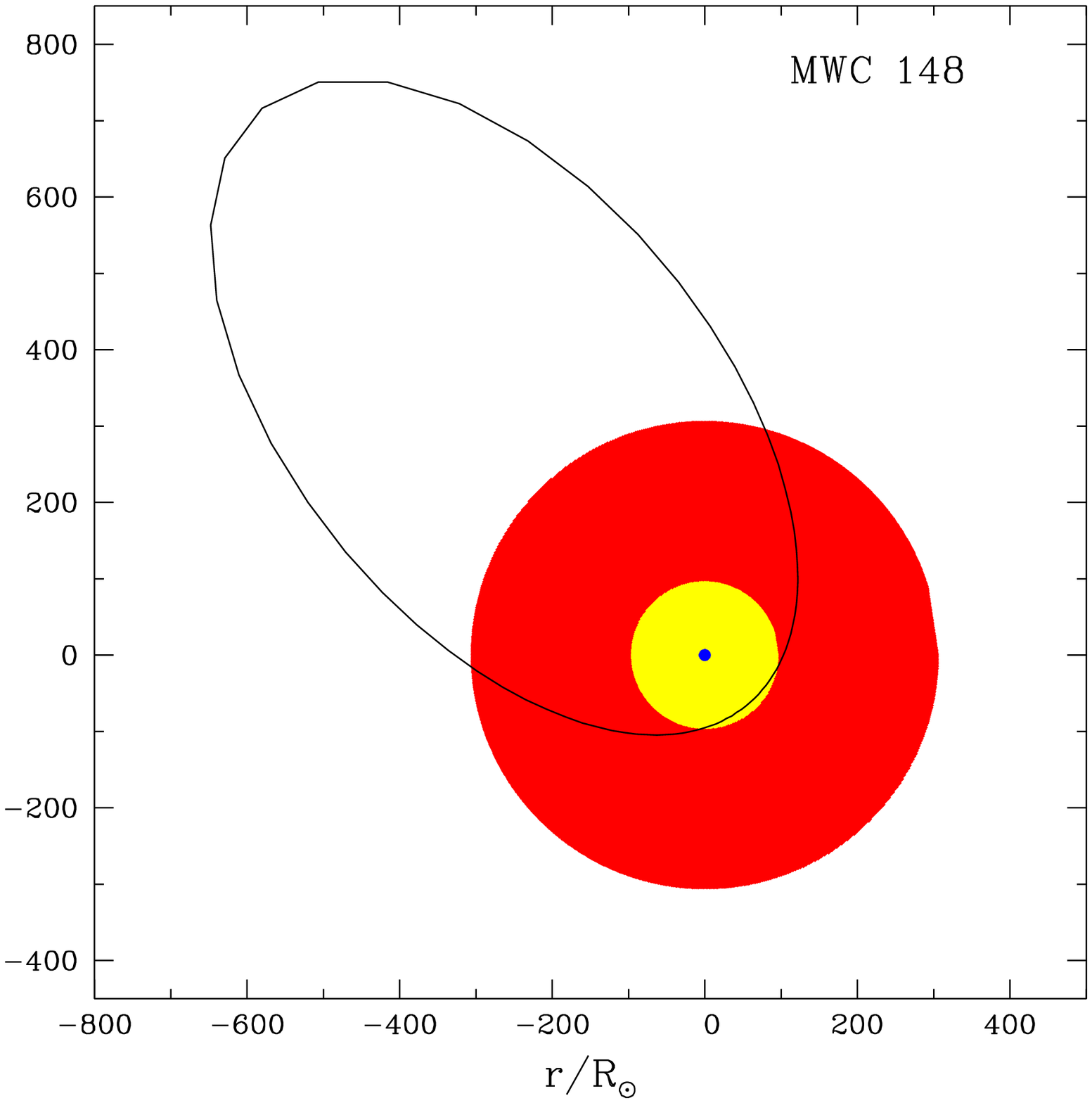}  
  \includegraphics{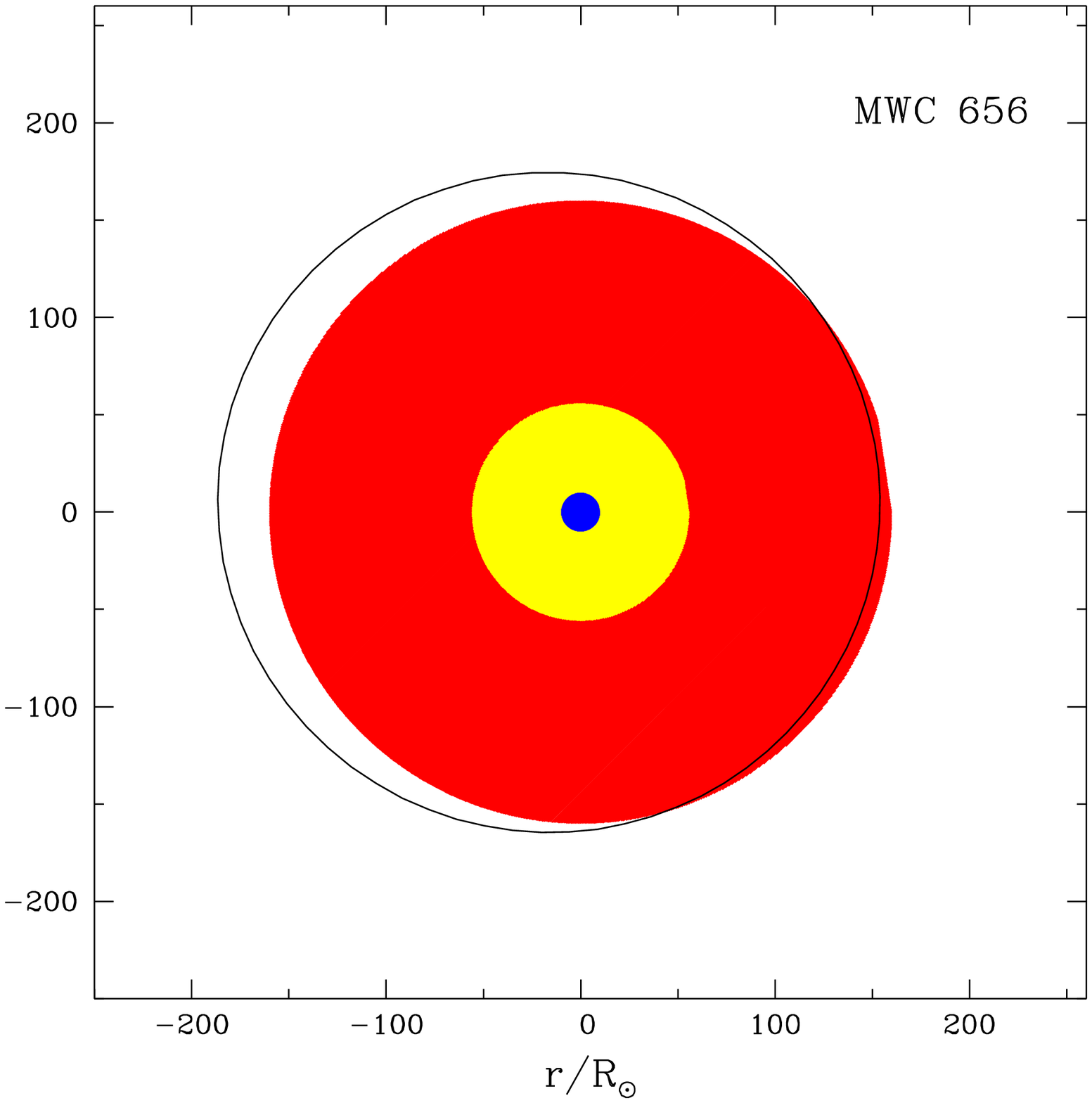}          
  \caption[]{Orbit of the compact object and circumstellar disc for \lsi, MWC~148, and MWC~656. 
Red indicates the disc size in $H\alpha$, yellow indicates the disc size in $H\beta$. The blue circle indicates the 
size of the mass donor.  We have three different cases: the orbit crossing the outer parts of the disc (\lsi), 
the orbit entering deeply in the disc at periastron (MWC~148), and the orbit at the disc border (MWC~656).}  
\label{f5.orbit} 
\vspace{7.5cm} 
\includegraphics{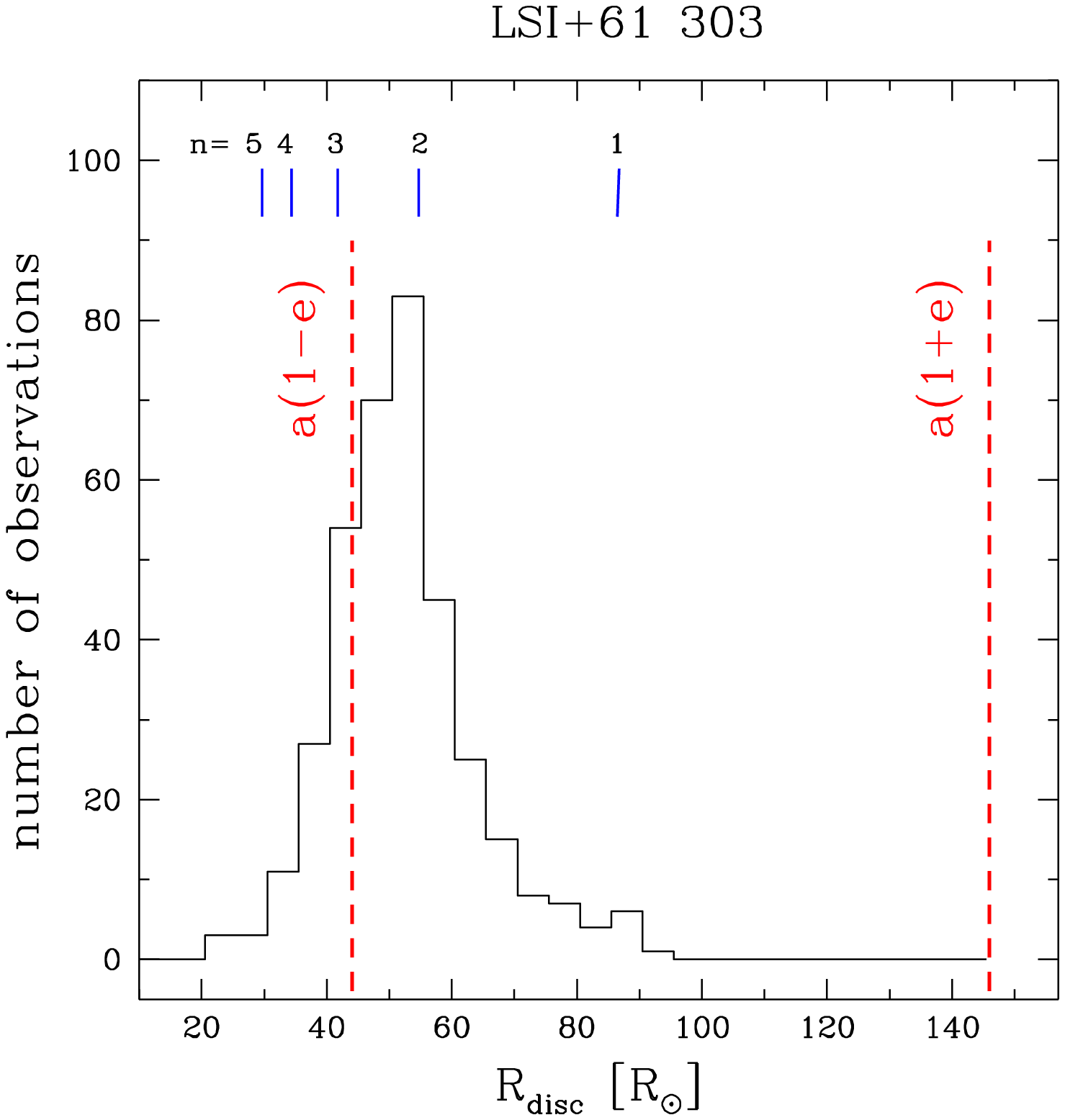} 
\includegraphics{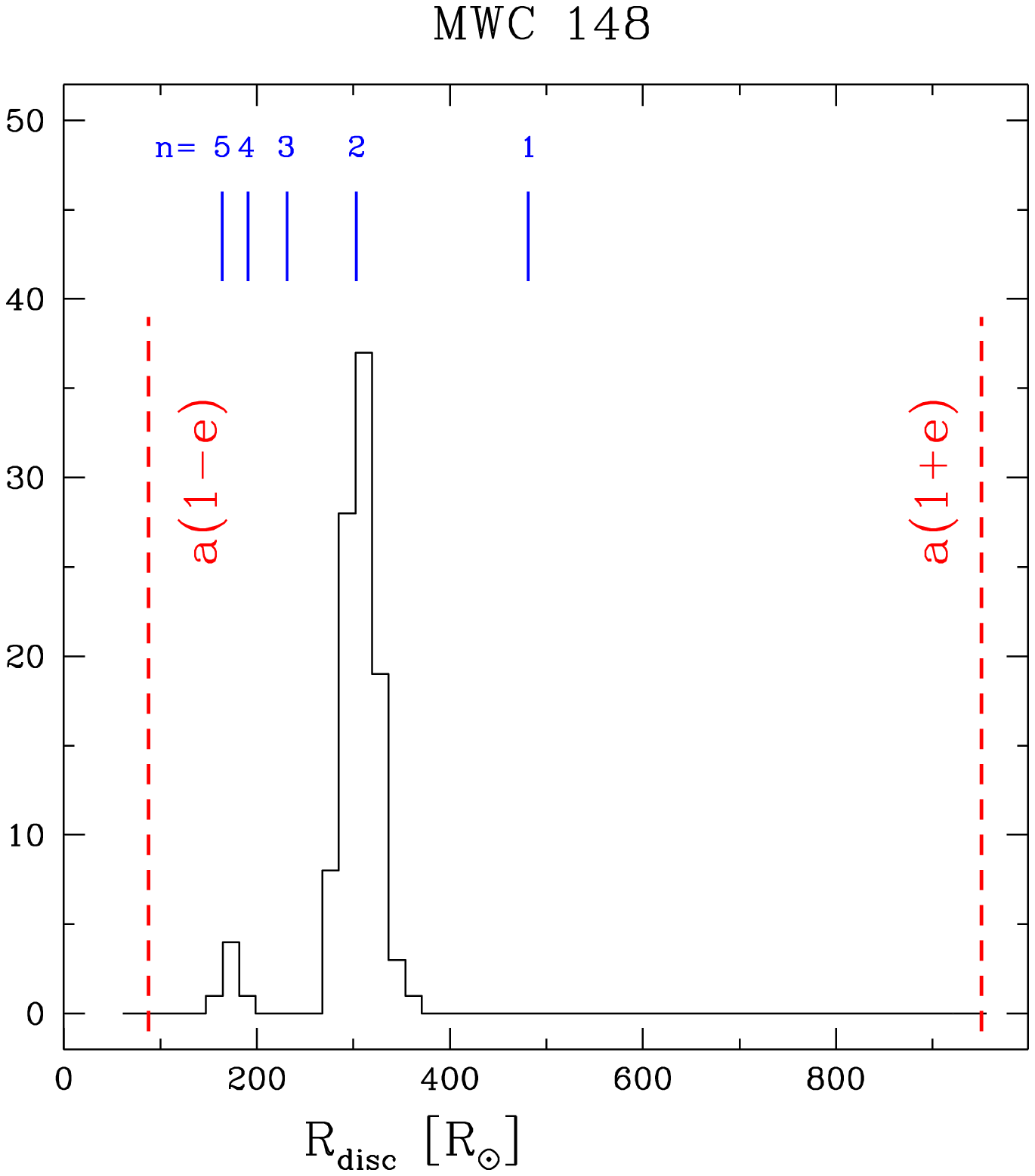} 
\includegraphics{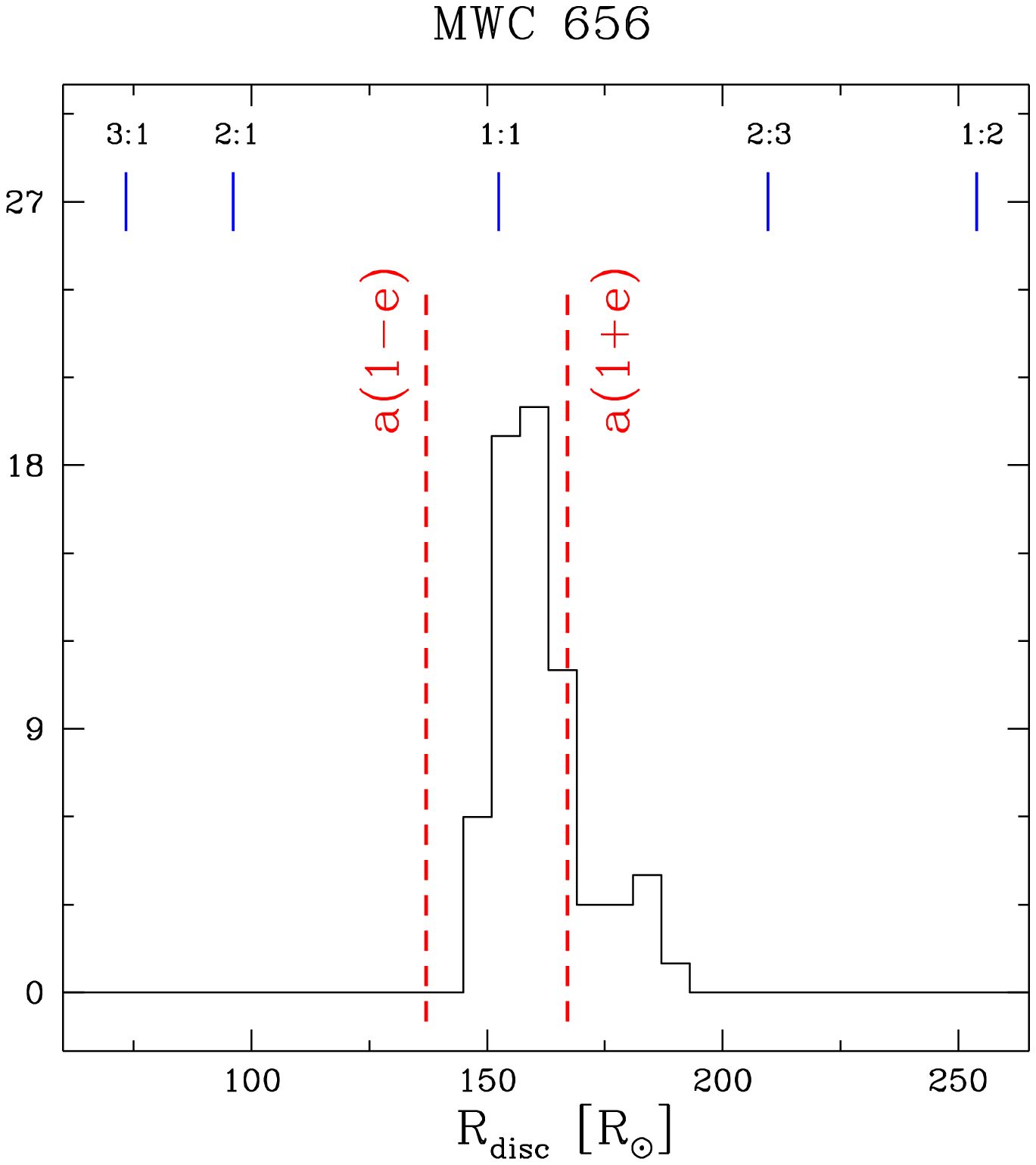} 
\caption[]{Histograms of the disc size. 
The vertical dashed (red) lines indicate the distance between the components  
at periastron and apastron, a(1-e) and a(1+e), respectively. 
The resonances n:m  are also indicated (in blue).  For \lsi\ and MWC~148 only those with m=1 are given. }
\label{f5.hist}  
\end{figure*}        

\subsection{Disc size and $W_\alpha$}
\label{RdWa}

The disc size and  $W_\alpha$ are  connected (see also Hanuschik 1989; Grundstrom \& Gies 2006). 
This simply expresses the fact that $R_{disc}$ grows as $W_\alpha$ becomes larger. 
In Fig.~\ref{f3.EW.dV}, we plot $\log \Delta V_\alpha / 2 $\vsi \   versus $\log W_\alpha$. In this figure
138 data points are plotted  for Be stars taken from 
Andrillat (1983), Hanuschik (1986), Hanuschik et al. (1988), Dachs et al. (1992), Slettebak et al. (1992), and Catanzaro (2013).
In this figure, the data for Be/$\gamma$-ray binaries are also plotted.  
The three Be/$\gamma$-ray binaries are inside the distribution of normal Be stars.

There is a  moderate to strong correlation between the variables with  Pearson correlation coefficient -0.63, 
Spearman's (rho) rank correlation 0.64, and  $p$-$value \sim 10^{-15}$. 
The dependence is of the form 
 \begin{equation}
     \; \;  \log \; (\Delta V_{\alpha}/2 v \sin i ) = -a \; \log W_\alpha + b, 
  \label{Han2}
 \end{equation}
and the slope is shallower  for stars with $ W_\alpha < 3$~\AA \  as noted by Hanuschik et al. (1988). 
For  120 data points in the interval $3 \le W_\alpha \le 50$~\AA, 
using a least-squares approximation
we calculate $a=0.592 \pm 0.030$ and  $b=0.165 \pm 0.036$.
This fit as well as the correlation coefficients
are calculated using only normal Be stars. 
Using  Eq. \ref{Huang} and Eq. \ref{Han2} we then obtain
 \begin{equation}
        \left( \frac {R_{disc}}{R_1}\;\right)^{-j} = 1.462 \; W_\alpha^{-0.592}.
  \label{Rd.W1}
  \end{equation}
Having in mind that (1) the discs of the Be stars are near Keplerian (Porter \& Rivinius 2003, Meilland et al. 2012); 
(2) the Be stars rotate at rates below the critical rate (e.g. Chauville et al. 2001),
and (3)  at higher optical depths the emission line peaks
are shifted towards lower velocities (Hummel \& Dachs 1992), 
we calculate the disc radius with the following  formula:
 \begin{equation}
        \frac {R_{disc}}{R_1}  = \; \epsilon \; 0.467 \; \; W_\alpha^{1.184}, 
  \label{Rd.W2}
  \end{equation}
where $\epsilon$ is a dimensionless parameter (see also Zamanov et al. 2013), for which  we adopt $\epsilon = 0.9 \pm 0.1$.
The disc sizes calculated with  Eq.~\ref{Rd.W2} are given in Table~\ref{tab.D} and are denoted with $(^c)$.
As can be seen, the values agree with those obtained with the conventional method. 
We estimate average values of the dimensionless quantity  
$R_{disc} / \epsilon R_1 = 8.7 \pm 1.9$ (for \lsi), 
$R_{disc} / \epsilon R_1 = 43 \pm 5$ (for MWC~148), and 
$R_{disc} / \epsilon R_1 = 18.0 \pm 1.1$  (for  MWC~656),  respectively.

%
%

\subsection{Disc truncation}
\label{disc.trunc}

The orbit of the compact object, the average size of $H\alpha$ disc, 
the average size of $H\beta$  disc, and the Be star are plotted in Fig. 3. The coordinates X and Y are in solar radii. 
 The histograms of $H\alpha$ disc size, calculated using Eq.~\ref{Rd.W2}, are plotted in Fig. 4.  
For \lsi,\ we use our new data and published data from  Paredes et al. (1994), Steele et al. (1996), Liu \& Yan (2005), 
Grundstrom et al. (2007), McSwain et al. (2010), and Zamanov et al. (1999, 2013). 
We use Rozhen and Liverpool Telescope spectra for MWC~148 and MWC~656. 
In all three stars the distribution of $R_{disc}$ values has a very well pronounced peak. 
The tendency for the disc emission fluxes to cluster at specified levels is
related to the truncation of the disc at specific disc radii by the orbiting compact object
(e.g. Coe et al. 2006). Okazaki \& Negueruela (2001) proposed
that these limiting radii are defined by the closest approach of the companion
in the high-eccentricity systems and by resonances between the orbital period and the disc gas rotational periods in 
the low-eccentricity systems. The resonance radii are given by
 \begin{equation} 
    {R_{n:m}^{3/2}} = \frac{m \; (G \: M_1)^{1/2}}{2 \: \pi} \:  \frac{P_{orb}}{n}, 
  \label{eq.resona}
  \end{equation} 
where $G$ is the gravitational constant,  $n$ is the integer number of disc gas rotational periods, and
$m$ is the integer number of orbital periods. 
The important resonances are not  only those with $n:1$, but can also be $n:m$ in general.

For \lsi\  (assuming $M_1 \approx 15$~\msun,  $M_2 \approx  1.4$~\msun, $e \approx  0.537$), 
we estimate the distances between the components  $a(1-e) \approx 44$~\rsun\ and 
$a(1+e)\approx 146$~\rsun\ for the periastron and apastron, respectively. 
As can be seen from Fig.~\ref{f5.hist}, the disc size is $R_{disc} \sim a(1-e)$ and it never 
goes close to  $a(1+e)$. The resonances that correspond to 
disc size are between  5:1  and 1:1, and the peak on the histogram 
corresponds to the 2:1 resonance. 

For MWC~148, with the currently available data ($M_1 \approx 15$~\msun,  $M_2 \approx  4$~\msun, $e \approx  0.83$), 
we estimate $a(1-e) \approx  88$~\rsun\  and 
$a(1+e)\approx 951$~\rsun\  for the periastron and apastron, respectively. 
In Fig.~\ref{f5.hist}, it is apparent  that  $a(1-e) < R_{disc} < a(1+e)$. 
The 2:1 resonance is the closest to the peak  of the distribution. 
We note in passing that the disc size in this star could have a bi-modal distribution
(a second peak with lower  intensity seems to emerge close to 4:1 resonance radius). 
 

For MWC~656 (assuming $M_1 \approx 9$~\msun,  $M_2 \approx  4$~\msun, $e \approx  0.1$), we estimate $a(1-e) \approx  137$~\rsun\ and 
$a(1+e)\approx 167$~\rsun\ for the periastron and apastron, respectively. 
In Fig.~\ref{f5.hist}, it is seen that the 1:1 resonance is very close to the peak  of the distribution 
and   $a(1-e)  \lesssim   R_{disc} \lesssim  a(1+e)$. 
The disc size rarely goes above  $a(1+e)$.


\section{Interstellar extinction: Estimates of $E(B-V)$ from interstellar lines}

There is a strong correlation between equivalent width of the diffuse interstellar bands (DIBs)
and reddening (Herbig 1975; Puspitarini et al. 2013). 
There is also a calibrated relation of reddening with the equivalent width of the 
interstellar line $KI \lambda7699$\AA\  (Munari \& Zwitter 1997).
Aiming to estimate the interstellar extinction towards our objects,
we measure equivalent widths of $KI \lambda7699$\AA\ and 
DIBs $\lambda6613, \lambda5780,  \lambda5797$. 


{\bf LSI+61$^0$303: } 
For this object, Hunt et al. (1994) use  $E(B-V)=0.93$ (Hutchings \& Crampton 1981).
Howarth (1983) using the 2200~\AA\ extinction bump obtained  $E(B-V)=0.75 \pm 0.1$.   
Steele et al. (1998) estimated  $E(B-V)=0.70 \pm 0.40$ from Na~I~D$_2$ and  $E(B-V)= 0.65 \pm 0.25$  from diffuse interstellar bands.
For \lsi,\ we measure 
$0.17 \le W(KI \lambda7699) \le 0.19$~\AA, 
$0.17 < W(DIB \lambda6613) < 0.19$~\AA, 
$0.34 < W(DIB \lambda5780) < 0.41$~\AA, 
$0.09 < W(DIB \lambda5797) < 0.16$~\AA, 
which following  Munari \& Zwitter (1997) and  Puspitarini et al. (2013) 
calibrations corresponds to  $E(B-V)=0.84 \pm 0.08$. 

{\bf MWC~148: } 
For this star, Friedemann (1992) estimated E(B-V)=0.85 from the 217 nm band. 
We measure  $0.14 < W(KI \lambda7699) < 0.20$~\AA,  
$0.13 < W(DIB \lambda6613) < 0.18$~\AA, 
$0.28 < W(DIB \lambda5780) < 0.34$~\AA, 
$0.13 < W(DIB \lambda5797) < 0.15$~\AA, 
which corresponds to a slightly lower value $E(B-V)=0.77 \pm 0.06$.

{\bf MWC~656: } 
For this star, Williams et al. (2010) gave a low value E(B-V)=0.02. 
Casares et al. (2014) estimated  E(B-V)=0.24. We measure 
$0.04 < W(DIB \lambda6613) < 0.06$~\AA,
$0.09 \le W(DIB \lambda5780) \le 0.10$~\AA, 
$0.03 < W(DIB\lambda5797) < 0.05$~\AA, 
and estimate $E(B-V)=0.25 \pm 0.02$. 


 \begin{figure}   
  \vspace{9.0cm}   
  \includegraphics{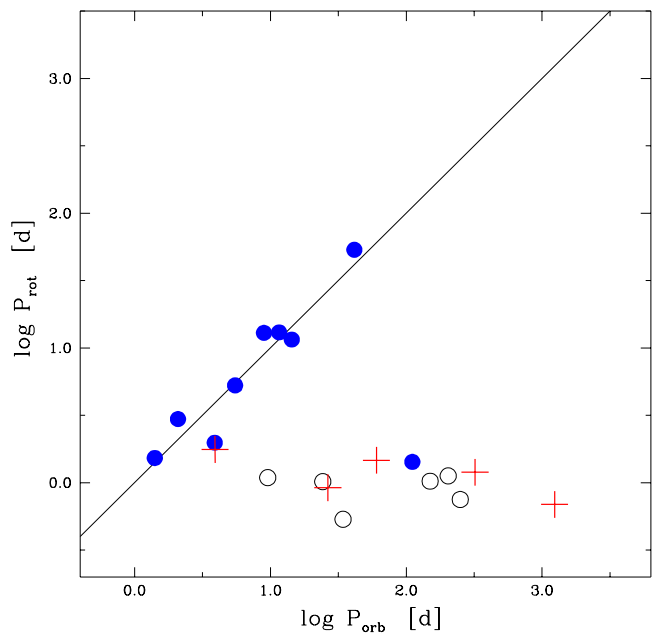}      
  \caption[]{Rotational period of the mass donor ($P_{rot}$) vs. $P_{orb}$. 
     Filled (blue) circles indicate high-mass X-ray binaries with giant/supergiant mass donor, open circles indicate Be/X-ray binaries, and red plusses indicate $\gamma$-ray binaries.  }  
\label{f2.Prot}      
\end{figure}         

\section{Rotational period of the mass donors}

In close binary systems the rotation of the companions of compact objects is accelerated by mass transfer and 
tidal forces  (e.g. Ablimit \& L\"u 2012). 
Adopting the parameters given in Sect.~\ref{sect.2},
we estimate the rotational periods of the mass donors $P_{rot} \approx 0.92$~d 
(for \lsi), $P_{rot} \approx 0.91$~d (MWC~148), and $P_{rot} \approx 0.86$~d (MWC~656).

For the $\gamma$-ray binary LS~5039, the mass donor is a O6.5V((f))  star with 
$R_{1} = 9.3 \pm 0.6 $~\rsun,  inclination $i \approx 24.9^0$,  \vsi $= 113 \pm 8$~\kms \  
(Casares et al.  2005).  We calculate $P_{rot} =  1.764$~d. 
Because of the short orbital period $P_{orb} = 3.91$~d, 
the rotation of the mass donor could be pseudo-synchronized with the orbital motion
(Casares et al.  2005). 


Radio observations of the pulsar in PSR B1259-63/LS 2883 
allow the orbital parameters 
to be precisely  established:  $P_{orb} = 1236.72$~d 
and eccentricity of e = 0.87 (Wang et al. 2004; Shannon et al. 2014). 
For the primary,  Negueruela et al. (2011) estimated 
$R_1 = 9.0 \pm 1.5$~\rsun,  \vsi \ $ = 260 \pm 15$~\kms, and  $i_{orb} \approx 23^0$, 
which give $P_{rot} =  0.689$~d. 


In Fig.~\ref{f2.Prot} we plot $P_{rot}$ versus $P_{orb}$ for a number of 
high-mass X-ray binaries (see also Stoyanov \& Zamanov 2009).   High-mass X-ray binaries with a giant/supergiant component, Be/X-ray binaries, and 
Be/$\gamma$-ray binaries are plotted with 
different symbols in this figure. 
The solid line represents synchronization ($P_{rot} = P_{orb}$). 
Among the five $\gamma$-ray binaries with known rotational velocity of the mass donor,
the  rotation of the mass donor is close to synchronization 
with the orbital motion in only one (LS~5039). The four 
Be/$\gamma$-ray binaries are not close to the line of synchronization.
They occupy the same region as the Be/X-ray binaries; in other words 
regarding the rotation of the mass donor
the Be/$\gamma$-ray binaries are similar to the Be/X-ray binaries.
In these, the tidal force of the compact object
acts to decelerate the rotation of the mass donor. 
The spin-down of the Be stars due to angular momentum transport from star to disc (Porter 1998)
is another source of  deceleration. 

In the well-known Be star $\gamma$~Cas, Robinson \& Smith (2000) found that the X-ray flux varied 
with a period P = 1.1 d, which they interpreted as the rotational period of the mass donor. 
A similar period is detected in photometric observations (Harmanec et al. 2000; Henry \& Smith 2012). 
This periodicity is probably due to the interaction between magnetic field of the Be star and its circumstellar disc
or the presence of some physical feature, such as a spot or cloud, co-rotating with the star.
The optical emission lines of MWC~148 are practically identical to those of $\gamma$~Cas.
All detected lines in the observed spectral range  (Balmer lines, HeI lines and FeII lines) 
have similar equivalent widths, intensities, profiles, and even a so-called wine-bottle structure noticeable in the $H\alpha$ line 
(see Fig.~1 in Zamanov et al. 2016).
Bearing in mind the above estimations and the curious similarities between: 
(i) the mean 20-60 keV X-ray luminosity of $\gamma$~Cas and \lsi\ (Shrader et al. 2015) 
  and 
(ii) the optical emission lines of $\gamma$~Cas and MWC~148, 
we consider that periodicity $\sim\!1$~day could be detectable 
in X-ray/optical bands in  the  Be/$\gamma$-ray binaries and could 
provide a direct measurement of the rotational period of the mass donor.

\section{Discussion}

The three Be/$\gamma$-ray binaries discussed here
have non-zero eccentricities and  misalignment 
between the spin axis of the star and the spin axis of the binary orbit could be possible 
(Martin et al. 2009). 

The inclination of the primary star in \lsi\ to the line of sight is probably   $ i_{Be} \sim  70^0$ (Zamanov et al. 2013).
Aragona et al. (2014) derived $a_1 \sin i_{orb} = 8.64 \pm 0.52$. 
Assuming $M_1 = 15$\msun, $M_2 = 1.4$~\msun, we estimate  $ i_{orb} \sim 67^0 - 73^0$. 
It appears that there is no significant deviation of the orbital plane from the equatorial plane of the Be star. 

The emission lines of MWC~148 are very similar to those of $\gamma$~Cas. 
The emission lines are most sensitive to the footpoint density and inclination angle (Hummel 1994). 
It means that in MWC~148 the Be star inclination should be similar to that of $\gamma$~Cas, 
for which the inclination is in the range $40^0 - 50^0$ (Clarke 1990, Quirrenbach  et al. 1997).
For MWC~148, Casares et al. (2012) estimated  $a_1 \sin i_{orb} = 77.6 \pm 25.9$, which for 
$M_1 = 15$~\msun\  and  a 4~\msun\ black hole gives  $ 45^0   \lesssim   i_{orb} \lesssim   65^0$.

For MWC~656,  Casares et al. (2014) give $M_1 \sin ^3 i_{orb} = 5.83 \pm 0.70$. 
Bearing in mind the range of  $8$~\msun~$\le M_1 \le 10$~\msun, this gives  $53^0  \lesssim i_{orb} \lesssim 59^0$. 
Inclination of  the Be star can be evaluated from the full width at zero intensity of the FeII lines 
$FWZI/2  \sin i = (G M_1/ R_1)^{1/2}$.
From  FWZI of FeII lines (Casares et al. 2012) and using $R_1= 9.5 - 10.0$~\rsun\   
we estimate $i_{Be} \approx 53 - 61^0$.  It appears that both planes are almost complanar. 

There are no signs of considerable deviation between the two planes. The opening half-angle
of the  Be stars' circumstellar disc are $\sim \! 10^0$ (Tycner et al. 2006; Cyr et al. 2015)
and it means that  the compact object is  practically orbiting in the plane of the circumstellar disc. 
The comparison between the orbit and circumstellar disc size (see Fig.~\ref{f5.orbit} and Sect.~\ref{disc.trunc})
shows that in these three objects we have three 
different situations:\begin{itemize}
\item  In \lsi\  the neutron star passes through the outer parts of the circumstellar disc at periastron, but
it does not enter deeply in  the disc; 
\item  In MWC~148 the compact object goes into the innermost parts of the disc
(passes through the innermost parts and penetrates deeply in  the disc) during the periastron passage;
\item  In MWC~656 the black hole is constantly accreting from the outer edge of the circumstellar disc. 
\end{itemize}
In MWC~656, this means that  the compact object (black hole) is 
at the disc border at all times and as a consequence it will have a higher and stable 
mass accretion rate along the entire orbit.  It seems to be a very clear case of 
disc truncation in which  the circumstellar disc is cut almost  exactly at the black hole orbit.

Because the $H\alpha$ peaks are connected with the outermost parts of the disc, 
the above three items  probably explain the observational findings:

1. In \lsi\ the $H\alpha$ emission line has a two-peak profile at all times (in all our spectra
in the period 1987 - 2015),because the circumstellar disc size is relatively small
and the neutron star passes only through the outermost parts of the circumstellar disc at periastron; 

2. In MWC~148, the big jumps in the $H\alpha$ parameters, 
$W_\alpha$, full width at half maximum and radial velocity (see Fig.~4 of Casares 2012)   
occur because 
the compact object enters (reaches) the inner parts of the disc during the periastron passage. 

3. In MWC 656 the double-peak profile is  not often visible because the black hole is 
 at the outer edge at all times and makes perturbations 
exactly in the places where the $H\alpha$ peaks are formed. 

It is worth noting that when the compact object causes large-scale perturbations, 
distorted profiles, such as those observed in  1A~535+262 (Moritani et al. 2011, 2013), will appear. 
If only a small portion of the outer disc is perturbed then it will appear 
in the central part of the emission line profile (e.g. in between the peaks or even filling the central dip)
because the outer parts of the disc produce the central part of the $H\alpha$ emission line profile. 
Similar additional emission is already detected in the $H\alpha$ spectra of 
\lsi\ (Paredes et al. 1990; Liu et al. 2000;  Zamanov \& Marti 2000).

 Gamma-ray emission has been repeatedly observed to be periodic in the
system LSI+61303 (Albert et al. 2009;  Saha et al. 2016)
and also very likely in MWC148, where its periodic X-ray flares are highly correlated with TeV emission
(Aliu et al. 2014).  A similar situation occurs in the case of  
LS~5039 (Aharonian et al. 2006), 1FGL J1018.6-20135856 (H.~E.~S.~S.~Collaboration et al. 2015), 
and PSR B1259-63 (H.E.S.S.~Collaboration et al. 2013),  
which has the longest orbital period observed in a $\gamma$-ray binary (about 3 yr). 
This is likely related to the very different physical conditions sampled by the compact companion 
as it revolves around the primary star in an eccentric orbit. 
Non-thermal emission (e.g. Dubus  2013) is produced from particles 
accelerated   at the shock between the wind of the pulsar 
and matter flowing out from the primary star (the polar wind as well as the disc). 

The X-ray and $\gamma$-ray light curve of MWC~148 shows two maxima at orbital phases 0.35 and 0.75 and minimum at 
apastron passage  (Acciari et al. 2009; Aliu et al. 2014). 
X-ray and $\gamma$-ray fluxes are correlated as mentioned above, in  agreement 
with  leptonic emission models, where relativistic electrons lose energy by synchrotron emission
and inverse Compton emission (Maier \& for the VERITAS Collaboration 2015).  The highly eccentric orbital geometry
sketched in the central panel of Fig.~\ref{f5.orbit} is also in good agreement with a periodic flaring system.


In the case of MWC~656, $\gamma$-rays have been detected occasionally (Williams et al. 2010), 
but so far never reaching the TeV energy domain. 
The main differences from previous sources are the fact that the compact object has been shown to be a black hole 
and its orbit is only moderately eccentric (Casares et al. 2014). 
Based on our spectroscopic observations, the size of the excretion disc is such that the black hole is accreting 
matter only from its lower density outer edges.  
As mentioned before, this implies that the accretion rate is stable  but at the same time low. 
Indeed, the quiescent X-ray emission level of the system is as weak as $\sim 10^{-8}$ Eddington luminosity 
(Munar-Adrover et al. 2014). 
Therefore, episodic $\gamma$-ray flares such as those detected by AGILE, likely require some enhancement of mass loss 
from the primary Be star or clumps in its circumstellar disc. 
We speculate here that the physical mechanism responsible for $\gamma$-ray emission in the MWC 656 black hole context 
could be related to the alternative microquasar-jet scenario also proposed for $\gamma$-ray binaries 
(see e.g. Romero et al. 2007). 
The fact that MWC~656 seems to adhere to the low-luminosity end of the X-ray/radio correlation for hard state 
compact jets also points in this direction (Dzib et al. 2015).

\section{Conclusions} 


From the spectroscopic observations of the three Be/$\gamma$-ray binaries we deduce 
that in \lsi\ the neutron star crosses the outer parts of the circumstellar disc at periastron,  
in MWC~148 the compact object passes deeply through the disc during the periastron passage, and 
in MWC~656 the black hole is  accreting from the outer parts of the circumstellar disc 
during the entire orbital cycle. The histograms in all three stars 
show that the disc size clusters at specific levels, indicating the circumstellar disc 
is truncated by the orbiting compact object. We estimate the interstellar extinction towards \lsi, MWC~148, and MWC~656.
The rotation of the mass donors is similar to that of the Be/X-ray binaries. 
We suggest that the three stars deserve to be searched for a periodicity of about $1.0$~day.

\begin{acknowledgements}
The authors are grateful to an anonymous referee for valuable comments and suggestions.
This work was partially supported by grant AYA2013-47447-C3-3-P from 
the Spanish Ministerio de Econom\'{\i}a y Competitividad (MINECO), and
by the Consejer\'{\i}a de Econom\'{\i}a, Innovaci\'on, Ciencia y Empleo of Junta de Andaluc\'{\i}a as 
research group FQM-322, as well as FEDER funds.

\end{acknowledgements}

\end{document}